\documentclass{llncs}
\usepackage{graphicx}
\usepackage{epstopdf}

\begin{document}

\title{High-efficiency three-party quantum key agreement protocol with quantum dense coding and Bell states}

\author{Wan-Ting He\inst{1}\and Jun Wang\inst{1}\and Tian-Tian Zhang\inst{1}\and Faris Alzahrani\inst{2}, Aatef  Hobiny\inst{2}\and Ahmed Alsaedi\inst{2}\and Tasawar Hayat\inst{2,3} \and
Fu-Guo Deng\inst{1,2}}

\authorrunning{F. Author et al.}

\institute{Department of Physics, Applied Optics Beijing Area Major Laboratory,
Beijing Normal University, Beijing 100875, China\\
\and
NAAM-Research Group, Department of Mathematics, Faculty of
Science, King Abdulaziz University,  Jeddah 21589,
Saudi Arabia \and
Department of Mathematics, Quaid-I-Azam University 45320,
Islamabad 44000, Pakistan}

%\address{Department of Physics, Applied Optics Beijing Area Major Laboratory,
%Beijing Normal University, Beijing 100875, China }

\maketitle

\begin{abstract}
We propose a high-efficiency three-party quantum key agreement protocol, by utilizing two-photon polarization-entangled Bell states and a few single-photon polarization states as the information carriers, and we use the quantum dense coding method to improve its efficiency. In this protocol, each participant performs one of four unitary operations to encode their sub-secret key on the passing photons which contain two parts, the first quantum qubits of Bell states and a small number of single-photon states. At the end of this protocol, based on very little information announced by other, all participants involved can deduce the same final shared key simultaneously. We analyze the security and the efficiency of this protocol, showing that it has a high efficiency and  can resist both outside attacks and inside attacks. As a consequence, our protocol is a secure and efficient three-party quantum key agreement protocol.

\end{abstract}

%
%\pacs{03.67.Lx, 42.50.Ex, 42.50.Pq}

\section{Introduction}

Quantum communication provides an unconditionally secure way for the transmission of  information, by exploiting the principles in quantum mechanics. In recent decades, this field gains much attention of researchers all over the world. There are many important branches of quantum communication for different tasks, such as quantum key distribution (QKD) \cite{bb84,QKD1,QKD2}, quantum secure direct communication (QSDC) \cite{QSDC1,QSDC2,QSDC3,QSDC4,QSDC5,QSDC6,QSDC8,QSDC9,QSDC10}, quantum secret sharing (QSS) \cite{QSS1}, and so on.
QKD supplies a secure way for two remote legitimate parties to create a private key  \cite{bb84,QKD1,QKD2}.
The parties in QKD  can detect the eavesdropper, say Eve, if she monitors
their quantum channel, by picking up a subset of their
outcomes obtained with two nonorthogonal measuring bases
to check eavesdropping. They can then discard the outcomes
when they find  Eve. Far different from QKD, QSDC gives an absolutely secure
approach for two parties to transmit their secret message directly, without producing the private key in advance. The first QSDC scheme was proposed by Long and Liu \cite{QSDC1} and it exploits the properties of Bell states and uses a block transmission
technique in 2002. Subsequently, Deng, Long, and Liu \cite{QSDC2} clarified the standard criterion for QSDC  explicitly in 2003, and they proposed an important two-step QSDC protocol by using the Einstein-Podolsky-Rosen photon pair blocks . Recently, these two-step QSDC protocols \cite{QSDC1,QSDC2} were  experimentally
implemented by two groups \cite{QSDC3,QSDC4}. In 2004,
Deng and Long  \cite{QSDC5} introduced the first QSDC protocol based on a
sequence of single photons, called quantum one-time pad scheme which has been recently experimentally
demonstrated by Hu et al. \cite{QSDC6} in a noisy environment
with frequency coding. In 2017, Wu et al. \cite{QSDC8}   proposed a high-capacity QSDC protocol with two-photon six-qubit hyperentangled states.
QSS is used to share a secret key among some agents of a boss
\cite{QSS1}, in which the agents can reconstruct the secret if and only if they
collaborate. QSS has a higher requirement than QKD because a potentially dishonest agent may injure the benefit of the boss. The inside attacks will increase largely the difficulty of the design of QSS schemes in practical applications.

Quantum key agreement (QKA)\cite{QKA1,QKA2,QKA3,QKA4,QKA5,QKA6,QKA7,QKA8,QKA9,QKA10,QKA11,QKA12,QKA13,QKA14,QKA15,QKA16,QKA17,QKA18,QKA19,QKA20,QKA21,QKA22,QKA23} is another interesting multiparty quantum communication. It is an extension of classical key agreement \cite{KA1}, by utilizing the principles of quantum mechanics, i.e., quantum no-cloning theorem, Heisenberg uncertainty principle, and the principle of quantum state superposition.
Ways of generating the shared keys are different in QKA and QKD protocols.
In a QKA protocol, a classical final shared key is derived by two (or more) parties as a function of information contributed by each of them \cite{QKA1}, it is generated by all the participants together. But in a QKD protocol, a shared key is decided by one participant and distributed to others.
A secure QKA protocol needs to satisfy four properties \cite{QKA2}: correctness, security, fairness, and privacy property. The correctness property requires that each participant should receive the correct secret key. The security property requires that outside eavesdroppers cannot obtain the shared key without being detected. The fairness property indicates that non-trivial subsets of the involved participants can determine the final shared key. And the privacy property requires that the sub-secret key of each participant should not be learned by any other \cite{QKA2}. Those properties make the QKA protocols more suitable for open insecure channels, and make it have the extensive application prospect in open network \cite{QKA3}.
Also, they increase the difficulty of designing a secure QKA protocol.
In 2004, Zhou et al.  \cite{QKA4} proposed a QKA protocol which  contains two users and utilizes the quantum teleportation technique. In 2010, Chong et al. \cite{QKA5} proposed a QKA protocol based on the BB84 protocol, utilizing a delayed measurement technique. Nevertheless, only two users are involved in the above protocols  \cite{QKA4,QKA5}.
In 2013, Shi et al.  \cite{QKA6} presented a multi-party QKA (MQKA) protocol   by using the entanglement swapping technique. In the same year, Liu et al. \cite{QKA7} pointed out that Shi et al.'s protocol is not a fair QKA protocol and then put forward another MQKA protocol with single particles. In 2014, a MQKA protocol using Bell state and Bell measurement was proposed by Shukla et al.  \cite{QKA8}. In 2016, two MQKA protocol were proposed by Sun et al. with cluster state  \cite{QKA9} and six-qubit states  \cite{QKA10} respectively. In the same year, Liu et al. \cite{QKA1} calculated the previous MQKA protocols into three categories: the complete-graph-type MQKA protocols \cite{QKA3,QKA7}, the circle-type MQKA protocols \cite{QKA2,QKA6,QKA8,QKA9,QKA10,QKA11}, and the tree-type MQKA protocol \cite{QKA12}. A circle-type protocol has a higher qubit efficiency than the complete-graph-type one. But in  Ref. \cite{QKA12}, they described an instructional mode of the attacks to the circle-type MQKA protocols, and claimed that those previous MQKA protocol cannot resist the collusive participant attacks (or called inside attacks). In 2016, Huang et al. \cite{QKA13} proposed a QKA protocol in travelling-mode utilizing single photons and rotation operations. In 2017, Cai et al. \cite{QKA14} presented another MQKA protocol with rotation operations. Subsequently, Cao et al. \cite{QKA15} proposed a MQKA protocol based on quantum search algorithm. Recently, Huang et al. \cite{QKA16} put forward a MQKA protocol with collective detection and rotation operations.
Up to now,  many other different QKA protocols have been proposed \cite{QKA17,QKA18,QKA19,QKA20,QKA21,QKA22,QKA23}.

In this paper, we   propose  a secure and efficient three-party QKA protocol with Bell states, following partially the idea in previous studies \cite{QKA2,QKA11}. In our protocol, the idea of quantum dense coding is used \cite{densecode}. Four unitary operations are used as encoding operations, which are performed by each participant on passing photons. The passing photons contains two parts, the first quantum qubits of Bell states and the single-photon states. Our protocol can prevent dishonest participants learning any useful information about other's sub-secret key. It can successfully resist both outside attacks and inside attacks. At the end of this protocol, all participants involved can deduce the final shared key simultaneously with very little information about the positions of those single photons announced by others. Moreover, by using the dense coding method, this protocol also possess a high efficiency.

% Note that the single-photon part accounts for a very small percentage.

%The rest of this article is organized as follows. In the next section, a new secure three-party QKA protocol is proposed. In  Section 3, we analyze the security of this protocol. And finally in Section 4, we make a short conclusion of our work.

\begin{figure}%[th]
\includegraphics[width=\textwidth]{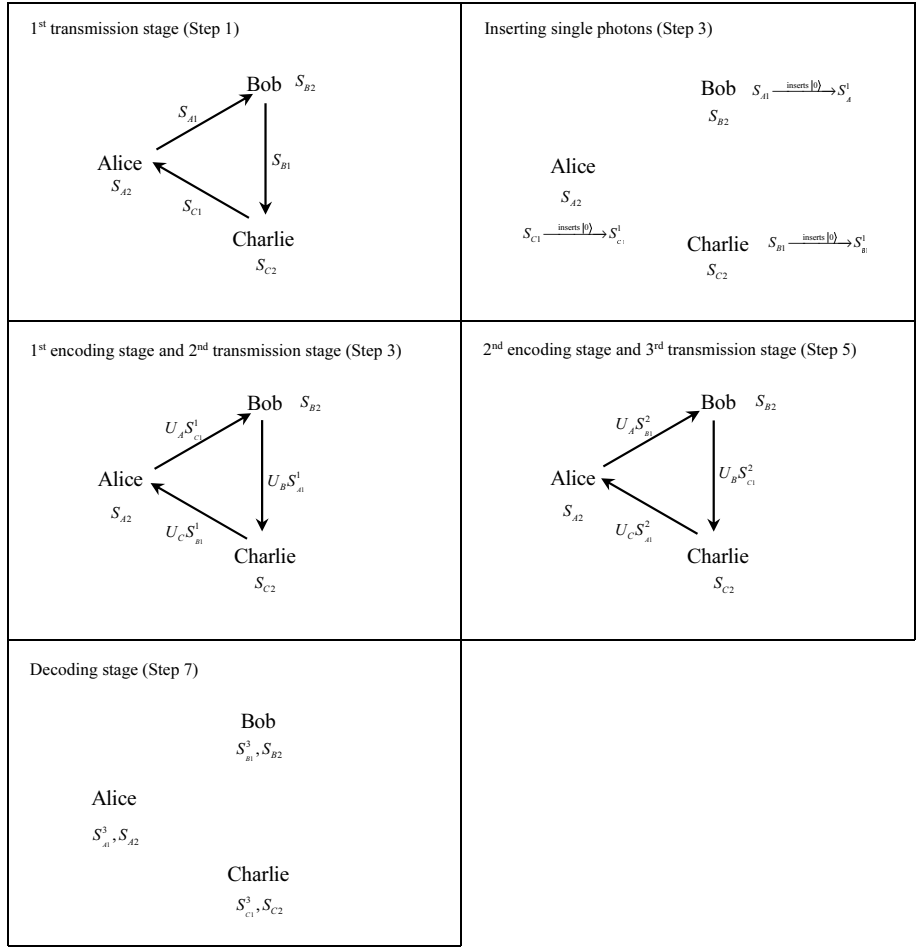}%\textwidth
\caption{The illustration of our three-party QKA protocol. $U_A$ ($U_B$, $U_C$) denotes the unitary operations performed by Alice (Bob, Charlie) according to his (her) secret bit string $K_A$ ($K_B$, $K_C$). The solid arrows present the sequence transmitted through the quantum channel with the block transmission technique \cite{QSDC1}.We ignore the eavesdropping check stages(Step 2, Step 4, Step 6) in this figure for conciseness.} \label{fig1}
\end{figure}

\section{High-efficiency three-party QKA protocol}

In this QKA protocol, three participants, say Alice, Bob, and Charlie, cooperate to establish a final shared key. The two-photon polarization-entangled Bell states and some single-photon polarization states will be used in our protocol. Four Bell states can be expressed as:
\begin{eqnarray}
\left| {{\phi ^ \pm}} \right\rangle &=& \frac{1}{{\sqrt 2 }}\left( {\left| {00} \right\rangle  \pm \left| {11} \right\rangle } \right),\nonumber\\              %  Eq.1
\left| {{\psi ^ \pm }} \right\rangle &=& \frac{1}{{\sqrt 2 }}\left( {\left| {01} \right\rangle  \pm \left| {10} \right\rangle } \right),
\end{eqnarray}
where ${\left| 0 \right\rangle }$ and ${\left| 1 \right\rangle }$ respectively present the horizontal and vertical polarization states of the single photon. They form a complete orthogonal basis, called Z basis. $\left|  +  \right\rangle  = \frac{1}{{\sqrt 2 }}\left( {\left| 0 \right\rangle  + \left| 1 \right\rangle } \right)$ and $\left|  -  \right\rangle  = \frac{1}{{\sqrt 2 }}\left( {\left| 0 \right\rangle  - \left| 1 \right\rangle } \right)$ form another complete orthogonal basis, called X basis.
We code the two-photon Bell state $\left| {{\phi ^ + }} \right\rangle$ as '00', $\left| {{\phi ^ - }} \right\rangle$ as '01', $\left| {{\psi ^ + }} \right\rangle$ as '10', $\left| {{\psi ^ - }} \right\rangle$ as '11'. Furthermore, we code the single-photon state ${\left| 0 \right\rangle }$ as '0', ${\left| 1 \right\rangle }$ as '1'.

In our protocol, according to the idea of quantum dense coding \cite{densecode}, we use four unitary operations $U_{00}=I$ , $U_{01}=\sigma _z$, $U_{10}=\sigma _x$ and $U_{11}=  i{\sigma _y}$ as the encoding operations. Here ${\sigma _z}$, ${\sigma _x}$, ${\sigma _y}$ are the Pauli matrices. If we choose to perform one of those local unitary operations on the first quantum bit of the two-photon system in the state $\left| {{\phi ^ \pm }} \right\rangle$ or $\left| {{\psi ^ \pm}} \right\rangle$, the transformation of those Bell states can be summarized in Table~\ref{tab1}. For single-photon states $\left|0\right\rangle$ and $\left|1\right\rangle$, their transformation by the those operations can be shown in Table ~\ref{tab2}.

%\makeatletter\def\@captype{table}\makeatother
\begin{table}%table 1
\caption{The transformation of four Bell states  $\left| {{\psi ^ \pm }} \right\rangle$ and $\left| {{\phi ^ \pm}} \right\rangle$ by the unitary operations $U_{00}$, $U_{01}$, $U_{10}$ and $U_{11}$ performed on the first quantum bits.}\label{tab1}
\setlength{\tabcolsep}{5mm}{
\begin{tabular}{|l|l|l|l|l|l|}
\hline
& ${U_{00}} \otimes I$ & ${U_{01}} \otimes I$ & ${U_{10}} \otimes I$ & ${U_{11}} \otimes I$ \\
\hline
$\left| {{\phi ^ \pm }} \right\rangle$ & $\left| {{\phi ^ \pm }} \right\rangle$ & $\left| {{\phi ^ \mp }} \right\rangle$ & $\left| {{\psi ^ \pm }} \right\rangle$ & $\left| {{\psi ^ \mp }} \right\rangle$\\

$\left| {{\psi ^ \pm }} \right\rangle$ & $\left| {{\psi ^ \pm }} \right\rangle$ & $\left| {{\psi ^ \mp }} \right\rangle$ & $\left| {{\phi ^ \pm }} \right\rangle$ & $\left| {{\phi ^ \mp }} \right\rangle$\\

\hline
\end{tabular}}
\end{table}

\begin{table}%table 2
\caption{The transformation of two single-photon states $\left|0\right\rangle$ and $\left|1\right\rangle$ by the unitary operations $U_{00}$, $U_{01}$, $U_{10}$ and $U_{11}$.}\label{tab2}
\setlength{\tabcolsep}{5mm}{
\begin{tabular}{|l|l|l|}
\hline
& $U_{00}\left( {{U_{01}}} \right)$ & $U_{10}\left( {{U_{11}}} \right)$ \\
\hline
$\left|0\right\rangle$ & $\left|0\right\rangle$ & $\left|1\right\rangle$\\
$\left|1\right\rangle$ & $\left|1\right\rangle$ & $\left|0\right\rangle$\\
\hline
\end{tabular}}
\end{table}

We assume that the classic channel is authenticated in our protocol, and we utilize the block transmission technique, which was first proposed by Long et al. \cite{QSDC1}, to ensure the security of transmission. Three participants  Alice, Bob and Charlie  want to establish a final shared key $K$.
Alice, Bob and Charlie first generate some random bit strings ${K_A}$, ${K_B}$ and ${K_C}$ as their secret bit strings (or called their sub-secret key), which can be expressed as:
\begin{eqnarray}                 %  Eq.2
{K_A} &=& ({a_1},{a_2}, \ldots ,{a_{n}}),\nonumber\\
{K_B} &=& ({b_1},{b_2}, \ldots ,{b_{n}}),\nonumber\\
{K_C} &=& ({c_1},{c_2}, \ldots ,{c_{n}}).
\end{eqnarray}
Here  ${a_i},{b_i},{c_i} \in \{ 00,01,10,11\} , i = 1,2, \ldots n$ . So the length of the each bit string is $2n$.\\
The illustration of our three-party QKA protocol are shown in Fig.~\ref{fig1}. They can be described in detail as follows.\\
\textbf{(Step 1)} Alice (Bob, Charlie) prepares $m$ two-photon systems $S_A$ ($S_B$,  $S_C$) in the maximally entangled state $\left| {{\phi ^ + }} \right\rangle$ and divides these particles into two ordered sequences, denoted as ${S_{A1}}$ and ${S_{A2}}$ (${S_{B1}}$ and ${S_{B2}}$, ${S_{C1}}$ and ${S_{C2}}$), respectively. Note that, these particles in ${S_{A1}}$ (${S_{B1}}$, ${S_{C1}}$) are the first qubits of the systems in $\left| {{\phi ^ + }} \right\rangle$, and these particles in ${S_{A2}}$ (${S_{B2}}$, ${S_{C2}}$) are the second qubits of the systems in $\left| {{\phi ^ + }} \right\rangle$. Alice (Bob, Charlie) prepares $kn$ single photons, which are randomly in the state $\{ \left| 0 \right\rangle ,\left| 1 \right\rangle ,\left|  +  \right\rangle ,\left|  -  \right\rangle\}$, as the decoy photons. Then, he (she) inserts randomly those decoy photons into the sequence ${S_{A1}}$ (${S_{B1}}$ , ${S_{C1}}$) and sends the sequence to Bob (Charlie, Alice) through the quantum channel.\\
\textbf{(Step 2)} After confirming that Bob (Charlie, Alice) has received the sequence, Alice (Bob, Charlie) announces the positions of the decoy photons and the corresponding basis ($Z$ basis or $X$ basis) for the measurement on each decoy photon. Bob (Charlie, Alice) picks out and measures the decoy photons according to the announcement of Alice (Bob, Charlie). They compare the measurement results with the initial states of the decoy photons. If the error rate exceed the threshold, they abort the protocol; otherwise, they continue their quantum communication to the next step. \\
Certainly, Bob (Charlie, Alice) should exploit the complex eavesdropping-checking process \cite{Trojan1,Trojan2}  to   prevent an eavesdropper, say Eve, to eavesdrop the quantum communication with Trojan horse attack in this step. As shown in Ref. \cite{Trojan2}, he should let each photon received pass through a filter with which only the wavelengths close to the operating one can be let in, which is used to filter  out Eve's invisible photons and avoid the invisible photon eavesdropping scheme with which Eve utilizes the fact that the single-photon detector is only sensitive to the photons with a special wavelength.  Moreover, Bob (Charlie, Alice) should use a photon number splitter (PNS: 50/50), which is used to divide each signal into two pieces for some of the decoy photons,  to defeat the delay-photon Trojan horse attack \cite{Trojan2}.\\
\textbf{(Step 3)} Bob (Charlie, Alice) randomly inserts $l$ single photons in the state $\left| 0 \right\rangle$ into $S_{A1}$ ($S_{B1}$, $S_{C1}$) to form a new sequence $S_{A1}^1$ ($S_{B1}^1$, $S_{C1}^1$). Note that, $m+l=n$, $n$ is the half length of the secret bit strings, and $l$ accounts for a very small percentage of n. The purpose of inserting single photons here is to resist the possible attacks from dishonest participants.
Then, he (she) performs the unitary operation ${U_{{b_i}}}$ (${U_{{c_i}}}$, ${U_{{a_i}}}$) on each photon in $S_{A1}^1$ ($S_{B1}^1$, $S_{C1}^1$) according to his (her) secret bit string ${K_B}$ (${K_C}$, ${K_A}$). For instance, if ${b_i}$ (${c_i}$, ${a_i}$)$=00$, Bob (Alice, Charlie) chooses ${U_{00}}$ to perform on the $i$th photon of $S_{A1}^1$ ($S_{B1}^1$, $S_{C1}^1$). After the local operations, we denote the new sequence as $S_{A1}^2$ ($S_{B1}^2$, $S_{C1}^2$).
Bob (Charlie, Alice) prepares $kn$ decoy photons which are randomly in the state $\{ \left| 0 \right\rangle ,\left| 1 \right\rangle ,\left|  +  \right\rangle ,\left|  -  \right\rangle\}$, and inserts them into the sequence $S_{A1}^2$ ($S_{B1}^2$, $S_{C1}^2$). Then, Bob (Charlie, Alice) sends the sequence to Charlie (Alice, Bob).\\
\textbf{(Step 4)} After confirming that Charlie (Alice, Bob) has received the sequence, they perform the second eavesdropping check. The checking method is same as that in Step 2. If the error rate  exceeds the threshold, they abort the protocol; otherwise, they continue the quantum communication to the next step. Also, Charlie (Alice, Bob) should exploit the complex eavesdropping-checking process \cite{Trojan1,Trojan2}  to   prevent  Eve  to eavesdrop the quantum communication with Trojan horse attack. \\
\textbf{(Step 5)} After picking out the decoy photons, Charlie (Alice, Bob) performs the unitary operation on each photon in $S_{A1}^2$ ($S_{B1}^2$, $S_{C1}^2$) according to his (her) secret bit string ${K_C}$ (${K_A}$, ${K_B}$) and form a new sequence $S_{A1}^3$ ($S_{B1}^3$, $S_{C1}^3$). The encoding method is same as that in Step 3. Then, Charlie (Alice, Bob) prepares $kn$ decoy photons and inserts them into $S_{A1}^3$ ($S_{B1}^3$, $S_{C1}^3$). After that, Charlie (Alice, Bob) sends it back to Alice (Bob, Charlie).\\
\textbf{(Step 6)} After confirming that Alice (Bob, Charlie) has received the sequence, Alice, Bob and Charlie publicly announce the positions of the single-photon states that he (she) has insert in Step 3 through the authenticated classic channel. In order to ensure the correctness of this protocol, Alice, Bob and Charlie randomly select a set of positions to check whether each participant has received the correct secret key at the end of this protocol.
After that, they perform the third eavesdropping check. The checking method is  same as that in Step 4. If the error rate  exceeds the threshold, they abort the protocol; otherwise, they continue the quantum communication to the next step.\\
\textbf{(Step 7)} After picking out the decoy photons, the remaining photons form the sequence $S_{A1}^3$ ($S_{B1}^3$, $S_{C1}^3$). Now, Alice (Bob, Charlie) has the sequences $S_{A1}^3$ and $S_{A2}$ ($S_{B1}^3$ and $S_{B2}$, $S_{C1}^3$ and $S_{C2}$). Note that the length of the sequence $S_{A2}$ ($S_{B2}$, $S_{C2}$) is $m$, and the length of the sequence $S_{A1}^3$ ($S_{B1}^3$, $S_{C1}^3$) is $n$. Then they come to the decoding stage. For these positions where the single-photon states have been insert, Alice (Bob, Charlie) measures the photons from sequence $S_{A1}^3$ ($S_{B1}^3$, $S_{C1}^3$) in Z basis, and records the bit string of the outcomes as ${M_a}$ (${M_b}$, ${M_c}$). For the other positions, Alice (Bob, Charlie) performs Bell measurement on the corresponding photon pairs from $S_{A1}^3$ and $S_{A2}$ ($S_{B1}^3$ and $S_{B2}$, $S_{C1}^3$ and $S_{C2}$), and records the bit string of the Bell state measurement results as ${M_A}$ (${M_B}$, ${M_C}$). Then, Alice (Bob, Charlie) computes ${K_A}^\prime  = {M_A}\parallel {M_a}$ (${K_B}^\prime  = {M_B}\parallel {M_b}$, ${K_C}^\prime  = {M_C}\parallel {M_c}$), where $\parallel$ presents connecting two strings. After that, Alice (Bob, Charlie) picks out the same positions, as the single-photon state in $S_{A1}^3$ ($S_{B1}^3$, $S_{C1}^3$), from the bit string ${K_A}$ (${K_B}$, ${K_C}$). He (She) keeps the first bits of the two bits and moves them to the end of bit sequence. After that, he (she) would get a new bit string ${K_A}^*$ (${K_B}^*$, ${K_C}^*$). Thus, the length of ${K_A}^*$ (${K_B}^*$, ${K_C}^*$) is same as ${K_A}^\prime$ (${K_B}^\prime$, ${K_C}^\prime$). And it is easy to verify that ${K_A}^* \oplus {K_A}^{\prime }={K_B}^*  \oplus {K_B}^{\prime }={K_C}^*  \oplus {K_C}^{\prime }$ according to Table~\ref{tab1} and Table~\ref{tab2}. Here $\oplus$ denotes the addition module 2.\\

Finally, for each participant, he (she) can get the same final shared key $K = {K_A}^* \oplus {K_A}^{\prime }={K_B}^*  \oplus {K_B}^{\prime }={K_C}^*  \oplus {K_C}^{\prime }$. Then, they check whether the final shared key from those positions chosen in Step 6 is consistent. If not, they they abort the protocol.\\

%In our protocol, each participant sends its sequence to another one simultaneously. So he (she) is both a sender and a receiver at the same time. Their transmission paths just form a circle which is similar to Yin et al.' s protocol \cite{QKA8}.

%\begin{figure*}[!htb]                    %Figure 1
%\centering
%\includegraphics[width=12 cm]{fig1.eps}
%\caption{}\label{fig1}
%\end{figure*}

\section{Security Analysis}

In this section, we analyze the security of our protocol. For QKA protocol, we not only need to consider the attacks from outside eavesdroppers, but also need to consider the inside attacks being done by dishonest participants. Hence, the security analysis of QKA protocols is more complex than that in QKD protocols.

\subsection{Outside Attack}
%\subsubsection{intercept-resend attack}
%\subsubsection{entangle-measure attack}
%\subsubsection{Trojan horse attack}

Suppose there is an eavesdropper Eve (not the legitimate participants) who wants to steal the final shared key without being detected by the  legitimate  participants. To achieve this goal, she must obtain participants' sub-secret key through some means. Eve mainly has these attack means: intercept-resend attack, measurement-resend attack, entangle-measure attack and Trojan horse attack. Therefore, if we want to demonstrate the security of our protocol against the outside attacks, we must prove that our protocol can resist all those four attack means. Without loss of generality, we take the situation that the sequence is generated by Alice, and sends to Bob and Charlie orderly for example to describe the security of our QKA protocol.

First, let us consider the intercept-resend strategy. Eve intercepts the photons at the end of Step 1,  Step 3 or Step 5, and replaces them with her own photons to get the sub-secret key of Bob or Charlie. But she cannot pass the eavesdropping check. Eve couldn't know the information of the decoy photons which are randomly in the state $\{ \left| 0 \right\rangle ,\left| 1 \right\rangle ,\left|  +  \right\rangle ,\left|  -  \right\rangle\}$. Therefore, the probability that she will be detected is ${(1-\left( {\frac{1}{2}} \right)^{kn})}$. If $kn$ is large enough, Eve will be detected with the probability approaching to 100\%.

Second, let us consider the measurement-resending attack. Eve intercepts the photons and measures them at the end of Step 1, Step 3 or Step 5, and then resends them to Bob or Charlie. However, Eve can't distinguish between the target photons and decoy photons before the eavesdropping check process. So She just randomly chooses the measurement bases. As a result, Eve introduces many errors in the eavesdropping stage and exposes herself. The probability that she exposes herself is ${(1-\left( {\frac{3}{4}} \right)^{kn})}$ which will approach  to 100\% when $kn$ is large enough.

Now, let us come to the entangle-measuring attack. Eve intercepts the photons at the end of Step 1, Step 3 or Step 5, and employs a unitary operation on the ancillary photons $\left| E \right\rangle $ and the photons she captured.

Suppose the unitary operation is ${U_E}$, one can have the relations for the eavesdropping as follows:
\begin{eqnarray}                 %  Eq.3
{U_E}: \; \left| 0 \right\rangle \left| E \right\rangle  \;\; \to \!\!\!\!&& \left| 0 \right\rangle \left| {{E_{00}}} \right\rangle  + \left| 1 \right\rangle \left| {{E_{01}}} \right\rangle, \nonumber\\
\left| 1 \right\rangle \left| E \right\rangle \;\; \to \!\!\!\!&&  \left| 0 \right\rangle \left| {{E_{10}}} \right\rangle  + \left| 1 \right\rangle \left| {{E_{11}}} \right\rangle.
\end{eqnarray}
Here $\left| {{E_{00}}} \right\rangle$, $\left| {{E_{01}}} \right\rangle$, $ \left| {{E_{10}}}\right\rangle$ and $\left| {{E_{11}}} \right\rangle$ are pure states determined by ${U_E}$. In this protocol, we use the decoy photons randomly in the state $\{ \left| 0 \right\rangle ,\left| 1 \right\rangle ,\left|  +  \right\rangle ,\left|  -  \right\rangle\}$ to prevent eavesdropping.
At the end of Step 1, Step 3 or Step 5, if Eve use the entangle-measure attack, the states $\left|  +  \right\rangle$ ,$\left|  -  \right\rangle\ $will have the relations:
\begin{eqnarray}
                %  Eq.4
{U_E}:\left| + \right\rangle \left| E \right\rangle \;\; \to \!\!\!\! && \frac{1}{{ 2 }}\left|  +  \right\rangle \left( {\left| {{E_{00}}} \right\rangle  + \left| {{E_{01}}} \right\rangle  + \left| {{E_{10}}} \right\rangle  + \left| {{E_{11}}} \right\rangle } \right)\nonumber\\
 &&+ \frac{1}{2}\left|  -  \right\rangle \left( {\left| {{E_{00}}} \right\rangle  - \left| {{E_{01}}} \right\rangle  + \left| {{E_{10}}} \right\rangle  - \left| {{E_{11}}} \right\rangle } \right),\nonumber\\
\left| - \right\rangle \left| E \right\rangle \;\; \to \!\!\!\! && \frac{1}{{ 2 }}\left|  +  \right\rangle \left( {\left| {{E_{00}}} \right\rangle  + \left| {{E_{01}}} \right\rangle  - \left| {{E_{10}}} \right\rangle  - \left| {{E_{11}}} \right\rangle } \right) \nonumber\\
 &&+ \frac{1}{2}\left|  -  \right\rangle \left( {\left| {{E_{00}}} \right\rangle  - \left| {{E_{01}}} \right\rangle  - \left| {{E_{10}}} \right\rangle  + \left| {{E_{11}}} \right\rangle } \right).
\end{eqnarray}
If Eve wants to introduce no error in the eavesdropping check in Step 2, Step 4 or Step 6, it must satisfy ${\left| {{E_{01}}} \right\rangle }={\left| {{E_{10}}} \right\rangle }=0$ and ${\left| {{E_{00}}} \right\rangle }={\left| {{E_{11}}} \right\rangle }$. If not, the probability of Eve being detected is ${(1-\left( {\frac{1}{2}} \right)^{kn})}$.

And if Eve uses the entangle-measure attack to entangle the ancillary photons with the first qubits of $\left| {{\phi ^ + }} \right\rangle$, and then one will have the relations:
\begin{eqnarray}                 %  Eq.5
{U_E}:\left| {{\phi ^ + }} \right\rangle \left| E \right\rangle \;\; \to \!\!\!\! && \frac{1}{{\sqrt 2 }}\left( {\left| 00 \right\rangle  \left| {{E_{00}}} \right\rangle  + \left| 10 \right\rangle  \left| {{E_{01}}} \right\rangle
+ \left| 01 \right\rangle  \left| {{E_{10}}} \right\rangle  + \left| 11 \right\rangle  \left| {{E_{11}}} \right\rangle } \right).
\end{eqnarray}
Hence, if Eve wants to introduce no error (not be detected by the legitimate participants), we will have  ${U_E}:\left| {{\phi ^ + }} \right\rangle \left| E \right\rangle  \to \left| {{\phi ^ + }} \right\rangle \left| {{E_{00}}} \right\rangle $. Her ancillary photons and the encoded photons are in product states. She cannot get any useful information from measuring the ancillary photons. And we will come to the same conclusion when Eve entangles the ancillary photons with other Bell states or single-photon states. Thus we can say that Eve cannot obtain the sub-secret keys without being detected if she takes the entangle-measuring attack.

Finally, our protocol may suffer from the Trojan attack, which take use of the invisible photons. The legitimate participants can use the wavelength quantum filters and  PNSs to resist this attack \cite{Trojan1,Trojan2}.%,Trojan3,Trojan4

In conclusion, outside eavesdroppers can not obtain the shared key without being detected. According to the definitions in Ref. \cite{QKA2}, our protocol can reach the security property of the QKA protocol.

\subsection{Inside Attack}

The inside attack can also be seen as a violation of the privacy and fairness property \cite{QKA2}. The inside attack is that one or more dishonest participants want to predetermine the final shared key without being detected. There are two cases, having only one dishonest participant, or having two dishonest participants. If this protocol is immune to two participants attack, it is surely immune to one participant attack. So we only need to consider the second situation.
Suppose that Alice and Charlie are two dishonest participants, they want to conclude to determine the final shared key. In our protocol, the transmission route forms a circle. Choosing another two dishonest participants finally returns to the same situation.

Alice and Charlie must first get the sub-secret key of Bob ${K_B}$ before the Step 6. Then they can choose a different unitary operations to preform on the photons and then send back to Bob, or they tell Bob a fake information about the positions of single photons to predetermine the final shared key. In this protocol, the only chance they can get ${K_B}$ before Step 6 is  measuring photons in the $S_{A1}^2$  which have been performed unitary operations by Bob and the photons in $S_{A2}$ which has been preserved in their hands in the Step 5. But they do not know the positions of single photons which are randomly inserted by Bob, they cannot distinguish which two photons from sequence $S_{A2}$ and $S_{A1}^2$ form an EPR pair. They cannot get any useful information of ${K_B}$ to predetermine the final shared key. Thus,  all participants involved can equally influence the final shared key. This protocol satisfies the fairness property. And we can see in this protocol, the sub-secret keys of each participant are kept secret during the protocol. Even two dishonest participants cannot get any useful information of the honest participant. Hence this protocol can reach the privacy property. Besides, in Step 6, each participant randomly selects a set of positions to check whether they can conclude the correct secret key at the end of the protocol, ensuring the correctness property of this protocol.

In reality, outside eavesdroppers or inside dishonest participants may hide their attack under the channel noises. The quantum bit error rate introduced by channel noise is about 2\%-8.9\% \cite{QKA14,QBER1,QBER2,QBER3,QBER4,QBER5}. But in our protocol, the error rate introduced by outside or inside attack is at least 25\%. That is, our protocol will be immune to both the outside and inside attacks. And it can reach the correctness, security, fairness, and privacy property which a secure QKA protocol need to satisfy.

\section{Discussion and summary}

Now, we analyze the efficiency of our protocol, and show that the proposed protocol is efficient. According to the definition proposed by Cabello \cite{efficiency}, the efficiency of the quantum protocol is:
\begin{eqnarray}                 %  Eq.6
\eta  = \frac{c}{{q + b}}.
\end{eqnarray}
Here, $c$ denotes the number of the secret bits (here in QKA protocol, $c$ denotes the length of the classical final shared key), $q$ refers to the number of qubits transmitted in the quantum channel, and $b$ is the number of classical bits exchanged for decoding the message. In our protocol, comparing with $n$, $l$ and $kn$ are small quantities. So the efficiency of our protocol $\eta \approx \frac{2}{3}$. Comparing with previous QKA protocols under this definition of protocol efficiency, i.e. protocols with single photons \cite{QKA2,QKA7,QKA14,QKA16}, protocols with Bell states \cite{QKA8,QKA11,QKA17}, our protocol possess a higher efficiency in the three-party cases. This is because of dense-coding method which enables one qubit to carry two bits of information being used in the protocol.
It should make sense that we use the method of inserting some single photons rather than using control strings and performing rotation operations \cite{QKA13,QKA14,QKA16} to prevent the inside attacks.
With one encoding operation being performed, participant in those protocols can only encode one bit information on one photon while he can encode two bits in our protocol. In order to decode the message, participants in Ref.\cite{QKA13,QKA14,QKA16} need to exchange their control strings which are as long as the final shared key. While in our protocol, they only need to exchange very little information about the positions of those single photons. Thus, our protocol is easier to be implemented with less operations performed and less classical bits exchanged under this three-party condition.

According to the classification standard described in Ref. \cite{QKA1}, our protocol is a circle-type QKA protocol. And our protocol is not sensitive to collusive attacks (or called inside attacks) according to the above security analysis.
Two-photon entangled states and single-photon states are used as the information carriers in our protocol, they can be prepared and controlled\cite{remote1,remote2,remote3,remote4,remote5} even at a great distance. And photon states are also of great use in other fields, such as quantum computation\cite{compute1,compute2,compute3,compute4} and quantum simulation\cite{simulate1,simulate2}.
We utilize the Bell measurement to deduce the final secret key. Four polarization-entangled Bell states can be completely distinguished by the complete Bell state analysis method \cite{CBA}. Hence, our protocol is feasible under the current technologies and it can be implemented in realistic devices. Certainly, in a practical application of this QKA protocol with a noisy environment, some useful methods should be exploited to depress the influence of noise, such as decoherence-free subspace \cite{dfs1,dfs4,dfs5,dfs6}, self-error-rejecting transmission \cite{Kalamidas,lixhapl,faith4,faith7}, error correction with ancillary qubits \cite{Yamamoto}, entanglement purification \cite{EPP1,EPP2,EPP4,EPP5,EPP6,EPP7,EPP8,EPP9,EPP10,EPP12,EPP13,EPP14,EPP15,EPP16,EPP17,EPP18}, and entanglement concentration \cite{ECP1,ECP2,ECP3,ECP4,ECP5,ECP6,ECP7,ECP8,ECP9}.

In summary, we have proposed a three-party quantum key agreement protocol with  Bell states and quantum dense-coding method. We have analyzed the security of our protocol, showing that it is immune to both outside and inside attacks. And it can achieve the the correctness, security, fairness, and privacy property at the same time. Further more, compared with other protocols under this three-party condition, it is also a high-efficiency protocol.

\section*{ACKNOWLEDGMENTS}

This work is supported by the National Natural Science Foundation of
China under Grant No. 11674033, No. 11474026, and No. 11505007.

% and 11474026.

%, and the Fundamental
%Research Funds for the Central Universities under Grant No.
%2015KJJCA01.

\end{document}